\documentclass[journal=cmatex, layout=twocolumn]{achemso}

\usepackage{graphicx}

\usepackage{caption}

\usepackage{subcaption}

\usepackage{achemso}

\author{Alfonso Sanchez-Soares}

\author{James C. Greer}

\email{jim.greer@tyndall.ie}

\affiliation[Tyndall National Institute, Cork, Ireland]{Tyndall National Institute, Dyke Parade, Lee Maltings, Cork, T12 R5CP, Ireland}

\title{A semimetal nanowire rectifier: balancing quantum confinement and surface electronegativity}

\begin{document}

\begin{abstract}

For semimetal nanowires with diameters smaller than a few tens of nanometers, a semimetal-to-semiconductor transition is observed as the emergence of an energy band gap resulting from quantum confinement. Quantum confinement in a semimetal results in either lifting of the degeneracy of the conduction and valence bands in a \emph{zero} gap semimetal, or shifting of bands with a \emph{negative} energy overlap to form conduction and valence bands. For semimetal nanowires with diameters below 10 nanometer, the magnitude of the band gap can become significantly larger than the thermal energy at room temperature resulting in a new class of semiconductors relevant for nanoelectronics with critical dimensions on the order of a few atomic lengths. The smaller a nanowire's diameter, the larger its surface-to-volume ratio thus leading to an increasing impact of surface chemistry on its electronic structure. Energy level shifts to states in the vicinity of the Fermi level due to the electronegativity of surface terminating species are shown to be comparable in magnitude to quantum confinement effects at nanowire diameters of a few nanometer; these two effects can be used to counteract one another leading to semimetallic behavior for nanowire cross sections at which the quantum confinement effect would otherwise dominate. Abruptly changing the surface terminating species along the length of a nanowire leads to an abrupt change in the surface electronegativity. This can result in the formation of a semimetal-semiconductor junction within a \emph{mono}material nanowire, without the need for impurity doping nor requiring the formation of a heterojunction. Using density functional theory in tandem with a Green's function approach to determine electronic structure and charge transport, respectively, current rectification is calculated for such a junction. Current rectification ratios of $O( 10^3-10^5)$ are predicted from bias voltages as low as 300 mV. It is concluded that rectification can be achieved at essentially molecular length scales with conventional voltage biasing, while rivalling the performance of macroscopic semiconductor diodes.

\end{abstract}

\vspace{5mm}


\vspace{5mm}

Molecular electronics aspires to produce diode, transistor, or logic gate functions in single molecules or small molecular assemblies. Diode behavior is the most fundamental of these as rectifying junctions can be arranged and combined with electrostatic gating to provide a transistor or switching function, and hence the realization of all Boolean operations, as well as signal amplification, modulation, and other basic operations that enable electronic circuits. Molecular diodes have been demonstrated with increasing success and significant progress in recent years using new schemes such as tunable coupling~\cite{Batra2013} and rectification ratios ($I_{ON}/I_{OFF}$) greater than 200 have been reported using electrochemical methods~\cite{Capozzi2015}. In general, operating voltages tend to be relatively high ($> 1$ V) and rectification ratios low ($< 10$) rendering single molecular diodes mostly unsuitable for nanoelectronic designs, and reproducible placement and electrode coupling remains a challenge. \\

\vspace{5mm}

In the following, a strategy is presented which realizes a Schottky junction in a \emph{mono}material using ideas derived from molecular electronics in the sense that chemistry is used to locally modify electron affinity in order to control current flows~\cite{Aviram1974}. Applying this approach, surface terminating groups of differing electronegativity are introduced and shown to create semimetallic and semiconducting regions in $\alpha$-tin nanowires (SnNWs). The semimetal $\alpha$-tin ($\alpha$-Sn) is used as an exemplar to explore the effects of quantum confinement and altering of surface electronegativity through use of a combination of density functional theory (DFT) for electronic structure and a Green's function method for studying charge transport across the junction. $\alpha$-Sn nanowires are selected as they display advantageous properties for nanoelectronics in terms of the band gaps that can be achieved at small nanowire diameters~\cite{Ansari2012}, and relative to other semimetal it is straightforward to passivate the nanowires without the introduction of surface defect states within a quantum confinement-induced band gap. \par

Although not stable in bulk form at room temperature~\cite{Paul1961}, the $\alpha$ phase (diamond crystal structure) in thin films can be stabilized~\cite{Farrow1981,Hochst1983,John1989,Cheong1991} at room temperature. For a hydrogen terminated SnNW with diameter of approximately 5 nm, a semimetal-to-semiconductor transition is expected to yield a band gap of the order of 100 meV, with the band gap energy increasing rapidly to the order of an electron-volt with reducing nanowire cross section~\cite{Ansari2012}. A hydrogen terminated $\alpha$-Sn nanowire is used as a convenient reference system against which the electron affinity for different surface terminating species and band gap shifts are compared. It should be noted that a SnNW diameter of $\sim$1 nm diameter is used in this study, however due to deficiencies in the approximations to the exchange and correlation potentials in DFT, the quantum confinement induced band gaps are significantly under-predicted, as verified through {\it GW}~\cite{Hedin1965} calculations for SnNW band gaps~\cite{ASS2016C}. Hence the effects as demonstrated in the following are expected to extend to SnNWs with diameters estimated to be in the range of 3 nm to 5 nm. \par

It is known that surface chemistry results in a significant shift to the band gaps of semiconductor thin films and nanowires for critical dimensions below 10 nm. For example, in silicon nanowires increasing the electron affinity of the surface terminating groups results in a red shift to the band gap energy~\cite{Leu2006,Nolan2007}. Just as surface chemistry can reduce the effects of quantum confinement in semiconductor NWs, it will be shown that increasing the electronegativity of surface terminating groups or surface bonding atoms to a semimetal NW competes against the quantum confinement induced band gap at small dimensions. \par

\begin{figure}

\includegraphics[scale=1.0]{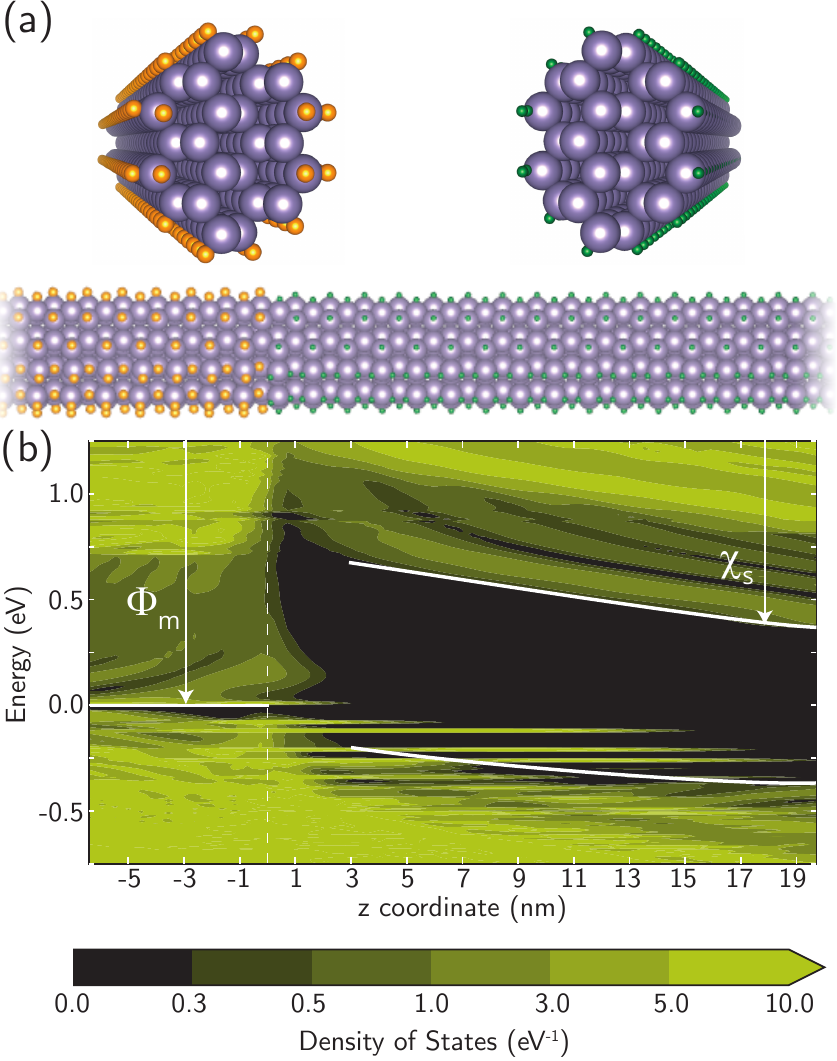}

\caption{Computed work function ($\Phi_m$), electron affinity ($\chi_s$), and bandgap ($\epsilon_G$) for 1 nm diameter $\langle 110 \rangle$ $\alpha$-Sn NWs passivated with fluorine and hydrogen. The larger electronegativity of fluorine is found to counteract the quantum-confinement induced bandgap observed for hydrogen terminated NWs.}\label{fig1}

\end{figure}

For the case of $\alpha$-tin NWs oriented along the $\langle110\rangle$ direction with fluorine termination, the band gap is reduced to the extent that the nanowire is nearly semimetallic with only a small energy gap of $\rm{\epsilon_g \sim k_BT}$ appearing at a diameter of $\sim$1 nanometer. As the quantum confinement effect reduces with increasing NW diameter, the fluorine terminated SnNWs are predicted to remain semimetallic at room temperature for diameters $>$ 1 nm. The effect of surface termination versus quantum confinement can be understood in terms of surface dipoles induced by the charge transfer at the surface of a NW to the terminating species. For large electron affinities, the terminating atoms or molecules draw charge out of the NW and the resulting surface negative charge can, as a first approximation, be thought of as a charge density distributed along the surface of a cylinder. The image potentials in the semimetal nanowire of opposite polarity may likewise be treated as a surface charge density but on the surface of an inner cylinder concentric to the outer cylindrical charge distribution. Hence the surface dipoles may be considered as two nested, hollow cylindrical charge distributions. The charges give rise to a potential difference between the two cylinders which can be expressed as 

\begin{equation} 
\Delta V = \frac{\rho_s}{2\pi\epsilon_0} \ln{\frac{r_o}{r_i}} \, ,
\end{equation}

where $\rho_s$ is the surface charge density, $r_o$ is the radius of the outer cylinder, ${r_i}$ the radius of the inner cylinder, and $2\pi\epsilon_0$ is the permittivity of free space. For two concentric cylinders, Gauss' law reveals there is no electric field within the region enclosed by both cylinders; the net effect of the cylinders is to produce a constant potential shift within the region enclosed by the inner cylinder. Hence, by analogy, the effect of increasing the magnitude of the surface dipole is to increase the potential shift at the center of the nanowire. A larger electronegativity of the terminating species increases the surface dipole leading to an increase in the magnitude of the electrostatic potential at the core region of the nanowire, or equivalently a lowering of the energy for an electron. The quantum confinement effect works to increase the energy of a confined electron and thus the partial cancellation between the two effects with increasing surface electronegativity.\\

\begin{figure} 

\includegraphics[scale=1.0]{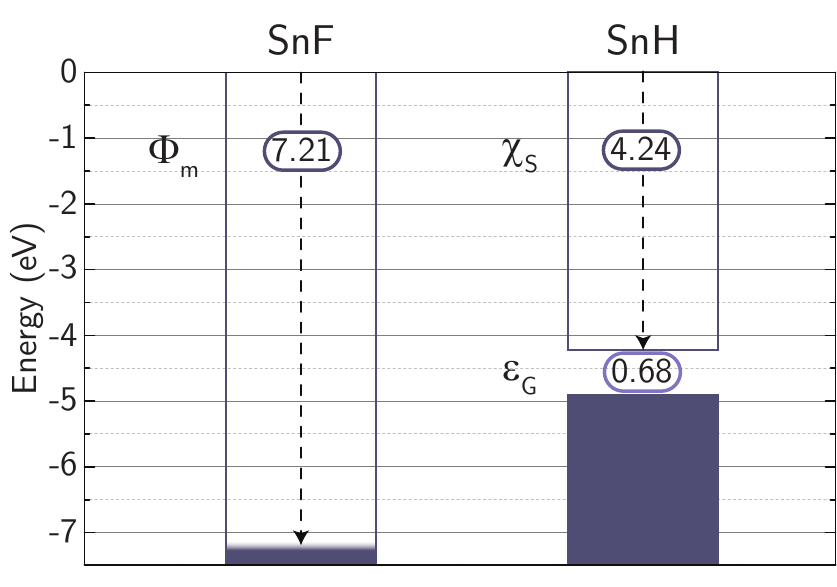}

\caption{A semimetal-semiconductor junction is realized by aburptly switching surface terminations from fluorine (left) to hydrogen (right) along the length of a $\langle 110 \rangle$ $\alpha$-Sn NW of diameter 1 nm. The contour plot shows the calculated local density of states (LDoS) across the junction, in which the work function of the semimetallic region ($\Phi_m$) and electron affinity of the semiconducting region ($\chi_s$) are indicated.}\label{fig2}

\end{figure}

The electronic structure of SnNWs is computed using DFT~\cite{Soler2002,QW}. The PBE~\cite{PERDEW96} formulation of the generalized gradient approximation (GGA) for the exchange-correlation functional has been employed in conjunction with norm-conserving pseudopotentials~\cite{Morrison93} and double-$\zeta$-polarized basis sets consisting of localized pseudo-atomic orbitals  (PAOs)~\cite{Ozaki03,Ozaki04}. The electronic structure for the 1 nm diameter $\langle$110$\rangle$ $\alpha$-Sn NWs are calculated with hydrogen and fluorine surface terminations with the results of the calculations presented in figure \ref{fig1} for band gap energies and electron affinities. For the hydrogen terminated SnNW, the energy band gap and electron affinity are found to be 0.68 eV and 4.24 eV, respectively. Upon replacing the surface terminating species by fluorine atoms, the electron affinity changes dramatically to 7.21 eV and the band gap nearly closes, being reduced to 30 meV. The large influence of the fluorine atoms' electron affinity is a reflection of the large surface-to-volume ratio with surface atoms accounting for $40\%$ of the chemical composition of the nanowire. The larger electronegativity of the fluorine atoms increases charge transfer at the NW surface relative to hydrogen termination with the correspondingly increased surface dipole offsetting the quantum confinement potential. The 30 meV band gap is of the order of $k_BT$ at room temperature and hence a significant population in the conduction band by thermally excited electrons results in semimetallic behavior.\\

To generate the atomistic model for the electronic structure and transport calculations, the total electronic energy is minimized with respect to atomic positions in a region of 5 nm around the junction until computed forces remain below $5 \times 10^{-2}$ eV/ \AA. Figure \ref{fig2}(a) shows the relaxed nanowire structure. In the left hand side of the figure the SnNW surface is terminated by fluorine atoms and in the right hand side the nanowire is terminated by hydrogen atoms. The largest structural rearrangements in the NW relative to the bulk diamond structure are caused by the surface terminating species with comparably little distortion to the NW core. In the region with hydrogen terminations the NW is semiconducting whereas the region terminated by fluorine atoms the local density of states (LDoS) indicates semimetallic behavior. Hence where the two regions meet, an abrupt semimetal-semiconductor junction is formed as shown in figure \ref{fig2}(b) where the LDoS is plotted in the absence of a voltage bias across the junction region. The vertical dashed line indicates the \emph{metallurgical junction} defined as the region across where the surface termination changes from fluorine to hydrogen. The resulting electronic structure clearly shows the characteristics of a Schottky junction with the upward band bending indicating charge transfer from the semiconducting region to the semimetallic side of the junction. On the other hand, the junction differs to the common engineering of Schottky diodes in that the semiconducting region induced by the hydrogen termination is intrinsic, hence the Fermi level aligns to mid gap in the semiconductor region at large distances away from the junction. The conduction and valence band bending upward in energy at the junction reflects the mismatch between the electron affinities in each region consistent with transfer of electrons from the semiconducting region to the semimetallic region. In this sense, the band bending of the junction resembles a metal/\emph{n}-type semiconductor junction. However, as the semiconducting region is an intrinsic semiconductor, the upward band bending of the conduction and valence band edges directly at the junction leads to the Fermi level aligning closer to the valence band edge. In this sense, the Fermi level alignment at the junction resembles a metal/\emph{p}-type semiconductor junction. These differences will lead to unconventional electron transport characteristics. A second characteristic of the junction is the presence of evanescent states or metal-induced gap states (MIGS)\cite{Bardeen1947} below the Fermi level and extending up to approximately 20 nm from the semimetallic region into the semiconducting region. Similar effects have been observed for an aluminum-silicon NW junction\cite{Landman2000} although in their study a short silicon nanowire is connected between two metal electrodes. Due to the Fermi level alignment near the valence band edge, the MIGS directly impact the current voltage (IV) characteristics across the junction, as will be seen from the charge transport calculations.\par

The alignment of the bands at the interface results in a Schottky barrier height (usually denoted as $\Phi_B$) of approximately 0.9 eV. The electron populations in the conduction band of the intrinsic semiconducting region and for the states in the semimetallic region above the Fermi level in the tail of the thermal distribution with energies near the top of the Schottky barrier are both negligible. These two conditions effectively rule out any significant electron flow between the semimetal and the semiconducting region's conduction band for typical voltage bias across the junction. \\

\begin{figure} [t]

\includegraphics[scale=1.0]{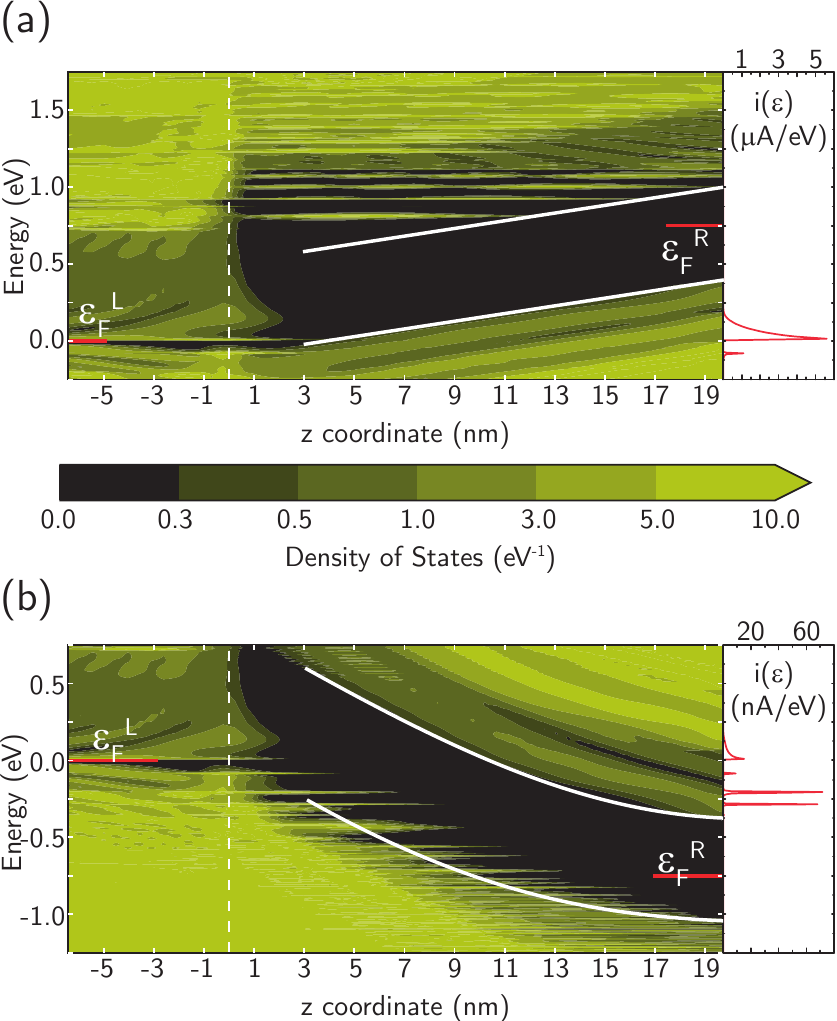}

\caption{Local density of states (LDoS) and energy-resolved current for a) Forward bias of +0.75 V. Note the quantization of states in the conduction band as a result of a triangular potential well-like profile near the junction; and b) Reverse bias of -0.75 V. Red segments on each side of the LDoS plots indicate the Fermi level alignments of each electrode. }\label{fig3}

\end{figure}

\begin{figure} [t]

\includegraphics[scale=1.0]{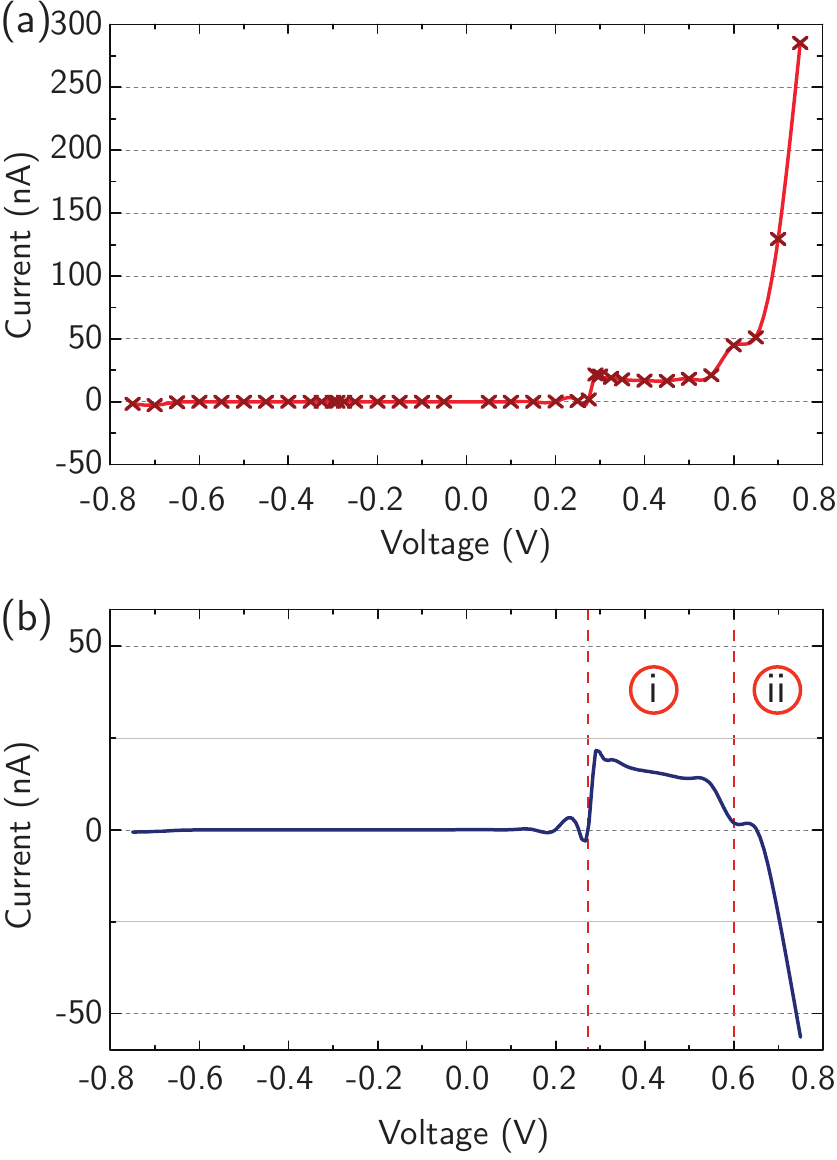}

\caption{a) Current-voltage characteristics of the junction at a temperature of 300K, and b) Difference between current-voltage characteristics at 300K and 0K.}\label{fig4}

\end{figure}

To investigate the charge flow across the junction with the application of an external voltage, the non-equilibrium Green's function (NEGF) formalism is applied~\cite{Datta2000,Brandbyge2002}. Figure \ref{fig3}(a) shows the LDoS with a forward bias of 0.75 V, defined as a positive voltage applied to the semimetallic region relative to the semiconducting region, is applied across the junction. For this voltage polarity, the electronic states in the semiconducting region shift towards higher energies relative to the semimetallic region. The built-in electric field arising from charge redistribution required to equilibrate the junction at zero bias is offset by the application of the voltage across the junction. The conduction mechanism for forward voltage bias arises from electron tunneling from the semiconducting region's valence band into the semimetallic region. For sufficiently high voltages, the MIGS at the semimetal's Fermi level mix with the semiconducting region's valence band states and the device becomes strongly \emph{ON}, as depicted in figure \ref{fig4}(a). This behavior is reflected in the energy-resolved current where a sharp peak is observed aligned near the Fermi level where these states extend into the valence band of the semiconducting region. This mechanism is distinct to the usual Schottky diode formed with an \emph{n}-type semiconductor under forward bias conditions for which electrons can more easily tunnel from a doped semiconductor's conduction band to the metal region as the applied voltage reduces the Schottky barrier height. \par

Reversing the voltage polarity across the junction results in the electronic structure depicted in figure \ref{fig3}(b): the semiconducting region becomes increasingly depleted of electrons as the reverse bias voltage is increased, leading to an increase in band bending near the junction relative to the zero bias band profile. In this case the quasi-Fermi level in the semiconducting region is lower relative to quasi-Fermi level in the semimetallic region. As shown in at the right in figure \ref{fig3}(b), electron transport occurs through band-to-band tunneling (BTBT) whereby electrons from the semimetallic region \emph{tunnel} through the semiconducting region's band gap and into the semiconducting region's conduction band. The largest contributions to the energy resolved current are seen to correspond to MIGS extending into the band gap within the voltage bias window. The MIGS reduce the distance required for BTBT in the structure. \par

Figure \ref{fig4}(a) presents the calculated IV characteristic for the junction for forward and reverse bias voltages self-consistently calculated within the NEGF formalism at a temperature $\rm{T=300 K}$. The asymmetry in the electron transport mechanisms with the polarity of the applied bias results in a two terminal rectifier or diode. The reverse bias leads to very low, BTBT limited transport. In the forward direction, the increasing electric field across the junction increases coupling between the valence band edge in the semiconducting region and states near the quasi-Fermi level in the semimetallic region leading to large current flows for applied voltages $V_a >$ 0.6 V. \par 

Notable in figure \ref{fig4}(a) is a structure above approximately $V_a$ =0.3 V extending to approximately $V_a$= 0.6 V. To explore this structure in the IV characteristic, transport across the junction at temperature $\rm{T=0K}$ was also computed and the plot in figure \ref{fig4}(b) gives the difference between the currents obtained at the two temperatures. There are two predominant regions where the currents differ significantly, in the range $0.3V < V_a < 0.6 V$ corresponding to the low current \emph{step-like} structure in the IV characteristic and for $V_a> 0.6V$ where the diode is strongly \emph{ON}. In figure \ref{fig1}(b), the equilibrium (zero applied voltage) electronic structure for the junction is shown. At approximately $V_a=0.3$ V, a flat band condition is achieved in the semiconducting region when the valence band edge aligns to the semimetal quasi-Fermi level. At T=300K, electrons in the semimetal region are readily excited above the Fermi level providing partially unoccupied states below $\epsilon_F^L$ into which the electrons in the semiconducting region's valence band maximum (VBM) can tunnel. The VBM provides an ideal conduction channel with transmission proportional to a one-dimensional density of states ($\propto 1/\sqrt{\epsilon}$), as is reflected in the IV characteristic. At $\rm{T=0K}$, there is no thermal excitation of electrons in the semimetallic region and all states below the Fermi level are blocked which results in lower currents as this channel is only open at higher temperatures -- Figure \ref{fig4}(b)(i).\par

At higher applied voltages, as depicted in figure \ref{fig3}(a), the states above the quasi-Fermi level in the semimetallic region are within the voltage bias window and electrons can tunnel from the valence band in the semiconducting region into these states which are unoccupied at T=0K. For T=300K, the states above the quasi-Fermi level are partially blocked due to their thermal occupations. Hence tunneling into lowest states within the voltage bias window is reduced at higher temperatures. Therefore the overall current is smaller at T=300K relative to T=0 K for $V_a > 0.6 V$ -- Figure \ref{fig4}(b)(ii). This property again reflects that the current transport mechanism for the surface modified semimetal junction is not the same as for  conventional Schottky diodes due to the differing band alignments.\par

These results demonstrate that Schottky barriers can be formed in a \emph{mono}material nanowire by exploiting the effects of quantum confinement and the impact of surface dipoles created by chemical terminating species with varying electronegativities. These two effects are of similar orders of magnitude in small cross section nanowires and can be manipulated to allow for semiconducting and semimetallic regions within a single nanowire or thin film leading to the possibility to form a semimetal-semiconductor junction, akin to Schottky barrier junctions formed in metal-semiconductor junctions in larger dimensions devices. The semiconducting region that is formed in a SnNW by quantum confinement results in an intrinsic material and hence the band alignments are different to conventional Schottky barrier diodes which rely on substitutional doping. However, the rectifying properties of the surface modified semimetal diode are comparable to the properties of conventional macroscopic diodes, yet are achievable at molecular scales and without the requirement for substutional doping in nanostructures. 

\section{Acknowledgements}

This work was funded by Science Foundation Ireland through a Principal Investigator award Grant No. 13/IA/1956. ASS was supported by an Irish Research Council graduate fellowship. We acknowledge additional support from Intel Corporation and QuantumWise. Atomistic visualizations were rendered using VESTA \cite{VESTA}.

\providecommand{\latin}[1]{#1}
\providecommand*\mcitethebibliography{\thebibliography}
\csname @ifundefined\endcsname{endmcitethebibliography}
  {\let\endmcitethebibliography\endthebibliography}{}

\end{document}